\documentclass[conference,10pt]{IEEEtran}
\IEEEoverridecommandlockouts
\usepackage{multirow}
\usepackage{threeparttable}

\usepackage{ulem}
\usepackage{makecell}
\usepackage{authblk}
\usepackage{diagbox}

\renewcommand{\baselinestretch}{1.0}

\usepackage{graphicx}
\usepackage{tabularx}
\usepackage{amsmath, amsthm}
\usepackage{amssymb}
\usepackage{mathtools}
\usepackage{comment}
\usepackage[subrefformat=parens,labelformat=parens]{subfig}
\usepackage{bm}
\usepackage{multirow}
\usepackage{threeparttable,booktabs}
\usepackage{tikz}
\usepackage[mathcal]{eucal}
\usepackage[]{algpseudocode}                               % algorithm package
\usepackage[ruled,lined,linesnumbered,noend]{algorithm2e}
\usepackage{filecontents}                                  % support to pgfplots
\usepackage{pgfplots}
\usepackage{pgfplotstable}
\pgfplotsset{compat=newest}
\usetikzlibrary{calc}
\usetikzlibrary{pgfplots.groupplots}
\theoremstyle{plain}

\newtheorem*{theorem*}{\textbf{Theorem}}

\theoremstyle{definition}

\newtheorem*{problem*}{\textbf{Problem}}

\usepackage[skip=1pt]{caption}            % set space between figure and caption
\graphicspath{{./figs/}}
\usepackage[customcolors]{hf-tikz}

\begin{document}

%\settopmatter{printacmref=false} % Removes citation information below abstract
%\renewcommand\footnotetextcopyrightpermission[1]{} % removes footnote with conference information in first column
%\pagestyle{plain} % removes running headers

%\IEEEoverridecommandlockouts
%\IEEEpubid{\makebox[\columnwidth]{978-1-6654-3274-0/21/\$31.00 ~\copyright2021 IEEE \hfill} \hspace{\columnsep}\makebox[\columnwidth]{ }}

% \copyrightyear{2023}
% \acmYear{2023}
% \setcopyright{acmcopyright}\acmConference[ICCAD '23]{IEEE/ACM International Conference on Computer-Aided Design}{October 30-November 3, 2022}{San Diego, CA, USA}
% \acmBooktitle{IEEE/ACM International Conference on Computer-Aided Design (ICCAD '22), October 30-November 3, 2022, San Diego, CA, USA}
% \acmPrice{15.00}
% \acmDOI{10.1145/3508352.3549398}
% \acmISBN{978-1-4503-9217-4/22/10}

% \settopmatter{printacmref=false} % Removes citation information below abstract

% \allowdisplaybreaks

\title{
\textbf{\huge READ: Reliability-Enhanced Accelerator Dataflow Optimization using Critical Input Pattern Reduction}}

% \author{Zuodong Zhang, Meng Li, Yibo Lin, Runsheng Wang, Ru Huang}

% \iffalse
\author[$1,3$]{Zuodong Zhang}
\author[$1$]{Renjie Wei}
\author[$2,1 \star$]{Meng Li \thanks{$\star$Corresponding author: meng.li@pku.edu.cn}}
\author[$1,3,4$]{Yibo Lin}
\author[$1,3,4$]{Runsheng Wang}
\author[$1,3,4$]{Ru Huang}
\affil[$1$]{\textit{School of Integrated Circuits, Peking University, Beijing, China}}
\affil[$2$]{\textit{Institute for Artificial Intelligence, Peking University, Beijing, China}}
\affil[$3$]{\textit{Institute of Electronic Design Automation, Peking University, Wuxi, China}}
\affil[$4$]{\textit{Beijing Advanced Innovation Center for Integrated Circuits, Beijing, China}}
% \fi

\maketitle

\begin{abstract}

With the rapid advancements of deep learning in recent years, hardware accelerators are continuously deployed in more and more safety-critical applications such as autonomous driving and robotics.
While the accelerators are usually fabricated with advanced technology nodes for high performance and energy efficiency, 
they are also more prone to timing errors under process, voltage, temperature, and aging (PVTA) variations.
By revisiting the physical sources of timing errors, we show that most of the timing errors in the accelerator are caused by a specific subset of input patterns, defined as critical input patterns.
To improve the timing error resilience of the accelerator, in this paper, we propose READ, a reliability-enhanced accelerator dataflow optimization technique that can effectively reduce timing errors.
READ reduces the occurrence of critical input patterns by exploring the optimal computing sequence when mapping a trained deep neural network to accelerators.
READ only changes the order of multiply-accumulate operations in a convolution, which introduces negligible hardware overhead and no impact on accuracy.
The experimental results on VGG and ResNet demonstrate on average 7.8$\times$ timing error rate (TER) reduction and up to 37.9$\times$ TER reduction for certain layers.
The results also show that READ enables the accelerator to maintain accuracy over a wide range of PVTA variations, making it a promising approach for robust deep-learning design.

\end{abstract}
\section{Introduction}
\label{sec:Introduction}

Deep neural networks (DNNs) have revolutionized different applications ranging from computer vision to natural language processing, 
and are widely deployed in data centers and edge devices. 
It can be foreseen that DNNs will be applied in more and more safety-critical applications like autonomous driving and robotics,
which typically require highly reliable computing to avoid catastrophic consequences.
Therefore, not only the model's robustness against various perturbations like adversarial noise, 
but also the robustness of the silicon-based accelerators to hardware faults need to be comprehensively investigated~\cite{tang2021robustart,liu2022fault,cores-dont-count,silent-data-corruption}. %\ml{Add a reference here?}

As the fabrication of DNN accelerators pushes toward nanoscale, 
the transient errors like timing errors that cannot be detected during manufacturing tests have become a more pronounced problem~\cite{mdat-20-Robust-ML}.
Timing errors due to the increased path delay usually occur under process, voltage, temperature, and aging (PVTA) variations.
Although DNN shows inherent error resilience at the algorithm level, timing errors are shown to cause significant accuracy degradation~\cite{DNN_Engine, thundervolt,jiaoxun_cnn_tolerance}.
This is because, on one hand, timing errors often occur in the most significant bit;
while on the other hand, the error will accumulate in each convolution operation and across the whole network \cite{jiaoxun_cnn_tolerance}.

% just propagating timing errors incur significant accuracy loss~\cite{DNN_Engine, thundervolt,jiaoxun_cnn_tolerance}.
% Because timing errors mostly occur in the most significant bit and the error will accumulate in a convolution operation, 
% hence it can significantly corrupt the model output.

Several timing error-resilient accelerator designs have been explored from the architecture level to the circuit level.
These works either utilize timing error detection and correction (TEDC) schemes to recover the correct value~\cite{thundervolt,EFFORT_journal,greentpu,DNN_Engine}, 
% or approximate computing to reduce path delay~\cite{aging-aware-quant,hussam_approximate_dnn}.
or algorithm-based fault tolerance (ABFT) techniques to check the correctness of computing~\cite{ABFT-cnn,ABFT-light}.
However, these approaches usually compromise network accuracy or introduce large hardware overhead.

\begin{figure}[tb]
    \centering
    \includegraphics[width=.48\textwidth]{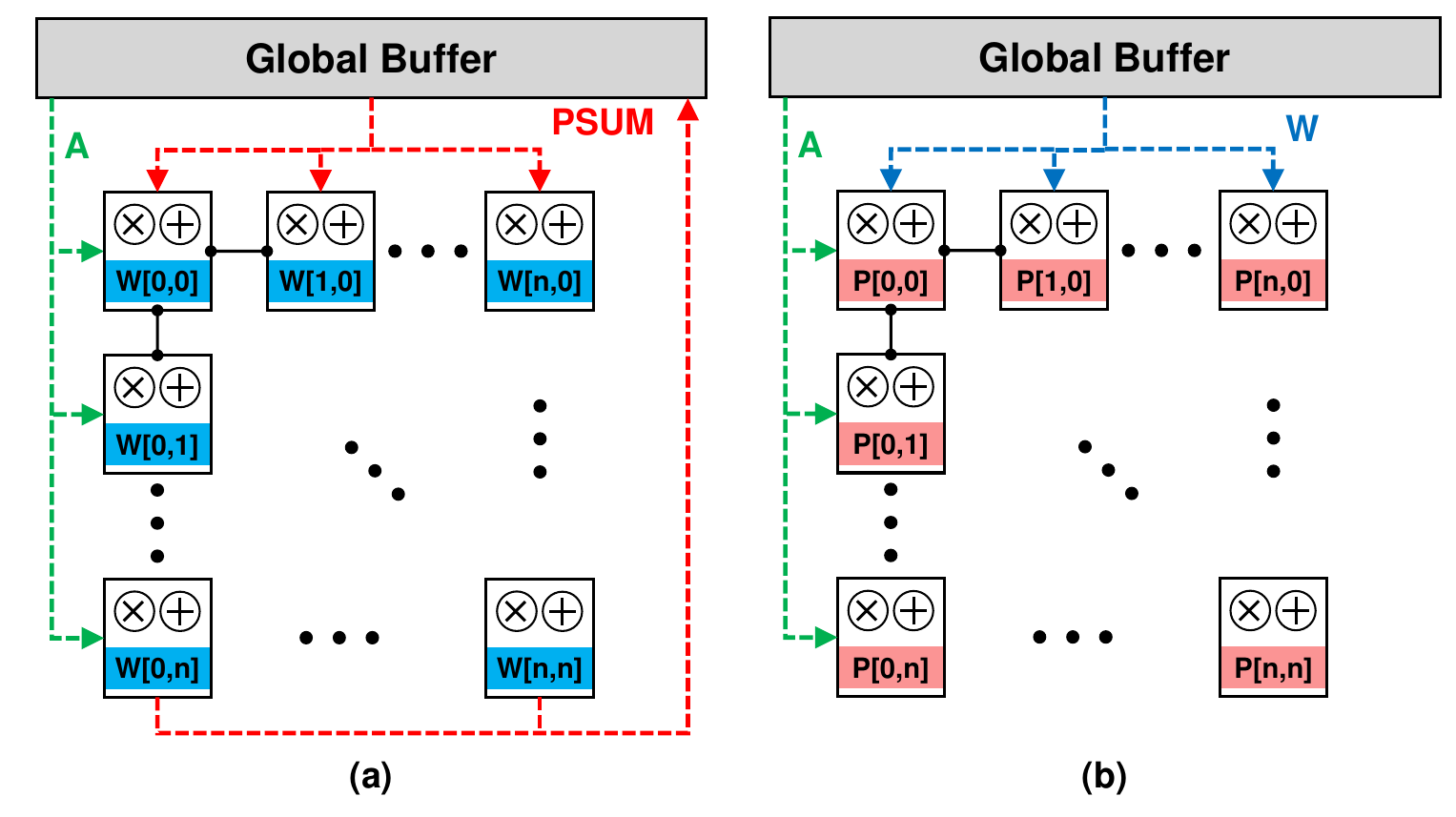}
    \caption{Dataflow for 2-D spatial accelerators. (a) Weight stationary. (b) Output stationary.}
    \label{fig:data-flow}
    % \vspace{-0.3cm}
\end{figure}

In this paper, we provide a promising solution to alleviate the timing error in DNN accelerators from a new perspective.
We propose READ, a reliability-enhanced accelerator dataflow optimization technique, 
which can improve the timing error resilience of DNN accelerators by exploring the optimal computing sequence.
To the best of our knowledge, READ is the first work that mitigates the accuracy loss of DNN accelerators due to timing errors by exploiting sequence optimization of dataflow,
and it is orthogonal to the previous TEDC and ABFT approaches.
The major contributions of this work are as follows:
\begin{enumerate}
    \item 
    % We observe that some input patterns are more prone to cause timing errors in multiply-and-accumulate (MAC) units. 
    We observe that timing errors are highly input pattern dependent and a subset of input patterns are significantly more vulnerable.
    The observation provides a new opportunity for timing error reduction by avoiding these critical patterns.
    
    \item We develop READ, a reliability-enhanced dataflow optimization technique to reduce timing errors in the multiply-accumulate (MAC) datapath.
    READ reduces the critical input patterns by weight re-ordering with negligible hardware overhead and no accuracy loss.
    
    \item We propose a clustering algorithm that divides the weight matrix into sub-matrices to improve the reordering flexibility and achieves better reordering results.
    
    \item The experimental results demonstrate on average 7.8$\times$ timing error rate (TER) reduction and up to 37.9$\times$ TER reduction for certain layers,
    which enables the accelerator to maintain accuracy over a wide range of PVTA variations.
    % makes the DNN accelerators more robust to PVTA variations. % \ml{Do we want to add the accuracy improvement as well here and in the abstract?}
    
\end{enumerate}

The rest of this paper is organized as follows. 
Section~\ref{sec:Background} introduces the background of DNN accelerators and the reliability concerns in DNN accelerators.
Section~\ref{sec:Motivation} introduces the motivation of the proposed optimization method.
Section~\ref{sec:Methodology} presents details of the algorithm. 
Section~\ref{sec:Results} demonstrated the experimental results of our algorithm. 
Finally, Section~\ref{sec:Conclusion} concludes the paper.

\section{Background}
\label{sec:Background}

\begin{table*}[t]
\centering
\captionof{table}{Features of the representative state-of-the-art timing error-resilient spatial accelerator design methods.}
\label{table:compare}
\begin{tabular}{l|c|c|c|c|c|c}
\hline \hline
\multicolumn{1}{c|}{\textbf{Methods}} & 
\multicolumn{1}{c|}{\textbf{Layer}} & 
\begin{tabular}[c]{@{}c@{}}\textbf{Scalable with} \\ \textbf{Technology}\end{tabular} &
\begin{tabular}[c]{@{}c@{}}\textbf{Accuracy} \\ \textbf{Loss} \end{tabular} &
\begin{tabular}[c]{@{}c@{}}\textbf{Hardware} \\ \textbf{Overhead}\end{tabular} &
\begin{tabular}[c]{@{}c@{}}\textbf{Throughput} \\ \textbf{Drop}\end{tabular} & 
\begin{tabular}[c]{@{}c@{}}\textbf{Design Effort}\end{tabular}  
\\ \hline \hline
Guardbanding    &circuit-layer    &no   &no     &High       &yes   &Low       \\  \hline
Sensitivity analysis~\cite{resilience-optimization-schorn,layer-wise-sensitivity}
                &algorithm-layer  &yes  &yes    &Negligible &no    &Medium    \\  \hline
ABFT~\cite{ABFT-cnn,ABFT-light}
                &algorithm-layer  &yes  &no     &Medium     &yes   &High      \\  \hline
Timing error detection~\cite{thundervolt,EFFORT-aspdac,DNN_Engine}
                &circuit-layer    &yes  &no     &High       &no    &Medium    \\  \hline
Timing error prediction~\cite{greentpu,jiatianyu-dnn}
                &circuit-layer    &yes  &yes    &Medium     &no    &High      \\ \hline
Ours            &dataflow         &yes  &no     &Negligible &no    &Low       \\ \hline \hline
\end{tabular}
\end{table*}

\subsection{ Spatial DNN Accelerators}

DNN accelerators aim to speed up the computationally expensive operations in DNN inference, e.g., computing the matrix multiplication $W \times A$. 
% The systolic array of MAC units, has been proposed as a promising direction~\cite{tpu_google}.
The 2-D spatial accelerators are a promising direction because of their capability to support parallel processing with minimal data movement~\cite{shidiannao,eyeriss,tpu_google}.
The 2-D spatial accelerators usually comprise the following major components: a two-dimensional arithmetic computing array, 
a network-on-chip for operand access and delivery, control blocks, and on-chip memory.

To reduce access to external memory, specialized processing dataflows are designed to enable data reuse across different process units.
The dataflow dictates what data gets read into the memory hierarchy and how data are propagated in the computing array.
Representative dataflows include weight stationary, and output stationary~\cite{accelerator_dataflow_review}.

Fig.~\ref{fig:data-flow} gives a schematic of weight stationary and output stationary.
The weight stationary dataflow~\cite{tpu_google,eyeriss} is designed to minimize the movement of the weights.
Each weight is read from the buffer into the register file (RF) of each PE and stays stationary,  
and the inputs and partial sums must move through the spatial array and global buffer. 
The output stationary dataflow is designed to minimize the partial sums movement. 
It keeps the accumulation of partial sums for the same output activation value local in the RF, 
stream the input activations across the PE array, and broadcast the weights to all PEs in the array~\cite{shidiannao}.

% \begin{figure}[tb]
%     \centering
%     \includegraphics[width=.48\textwidth]{fig/fig_add_dataflow.pdf}
%     \caption{Dataflow for 2-D spatial accelerators. (a) Weight stationary. (b) Output stationary.}
%     \label{fig:data-flow}
%     % \vspace{-0.3cm}
% \end{figure}

If the size of matrix-matrix multiplication is larger than the computing array, 
the total weight matrix will be tiled into sub-matrices and be performed in several blocks.

%%%%%%%%%%%%%%%%%%%%%%%%%%%%%%%%%%%%%%%%%%%%%%%%%%%%%%%%%%%%%%%%%%%%%%%%%%%%%%
\subsection{Reliability Concerns in Accelerator}
The transient errors in silicon-based accelerators can be divided into two categories, depending on where they occur.
The first is the soft errors that occur in the memory or on-chip buffers, and the second category is the timing errors that occur in the processing element (PE) \cite{thundervolt}.
The memory and on-chip buffers are usually protected with the error correction code (ECC), 
therefore, we mainly focus on timing errors.
The timing path of accelerators can be roughly divided into the timing path inside the computing array 
and the timing path between the computing array and the local cache for data exchange. 
Among them, the timing path inside the computing core is more critical.
Previous works~\cite{DNN_Engine,ares_dnn,jiaoxun_cnn_tolerance} have also demonstrated that the accuracy of DNN can be very sensitive to timing errors in computing array because of the following reasons:
1) the timing errors usually occur at the most significant bit, which is catastrophic for the computing results;
2) timing errors may occur in every single MAC operation, while thousands of MAC operations are required to compute a single output activation in the convolution layer.
The bit error rate (BER) of the output activation can be estimated by the TER of MAC as below:
% \vspace{-0.3cm}
\begin{equation}
    \label{BER_cal}
    BER(i) = 1 - \prod^N (1 - TER(i))
    % \vspace{-0.2cm}
\end{equation}

where $N$ is the number of MAC operations required for an output activation.
Even if the TER of a single MAC operation is relatively small, the BER of the output activations can be large.
Therefore, the inference accuracy will decrease rapidly with the increase of timing errors.

At the same time, accelerators fabricated with advanced technology nodes are more prone to timing errors under PVTA variations~\cite{review_huang}.
Thus the conventional guardbanding will not only result in excessive performance loss, 
but also cannot guarantee the requirements in safety-critical scenarios.
All these reliability concerns aggregate the uncertainty of the deep learning processing and hinder the deployment of deep learning in safety-critical applications.

\subsection{Related Works}
\label{sec:related-work}
To improve the timing error resilience of accelerators, a number of approaches from algorithm-layer to circuit-layer have been proposed.
We summarize the features of the representative state-of-the-art reliability-enhanced techniques in Table~\ref{table:compare}.
The algorithm-layer techniques utilize the NN's inherent redundancy and tolerance for errors, 
and common methods include sensitivity analysis, fault/noise-tolerant training, and model architecture design. 
% Choi et al. proposed to evaluate filter/channel-level weight sensitivities and assign sensitive weights to robust PE to improve error resilience~\cite{sensitivity_analysis_dnn}.
Libano et al. conduct layer-wise sensitivity analysis and protect the vulnerable layer~\cite{layer-wise-sensitivity}.
Schorn et al. propose a training strategy to adjust neuron sensitivities~\cite{resilience-optimization-schorn}, 
which argues that achieving a homogeneous resilience distribution can help DNN robustness.
% Liu et al. proposed to use error-correcting output codes (ECOC) to tolerate variations and faults~\cite{ECOC-dac}.
Zhao et al. adopt the algorithm-based fault tolerance (ABFT) technique of error-tolerant matrix-matrix
multiplication for convolution operations to obtain the error detection/correction capability~\cite{ABFT-cnn}, 
and Filippas et al. further proposed a lightweight ABFT implementation~\cite{ABFT-light}.
% Our work is a post-training optimization technique, so the model-layer techniques are orthogonal to our work.

Circuit-layer techniques usually utilize timing error detection (using Razor flip-flops~\cite{razor}, for instance) and correction approaches.
Zhang et al. introduced a new timing speculation methodology for accelerators, 
in which the MAC unit encountering a timing error, will steal an execution cycle from its downstream MAC to recover the correct value~\cite{thundervolt}.
Gundi et al. further improved the correction scheme~\cite{EFFORT-aspdac}. 
Whatmough et al. proposed a scheme that utilizes lightweight Razor flip-flops and circuit-layer timing borrowing to tolerate the timing errors~\cite{DNN_Engine}.

To the best of our knowledge, READ is the first one that exploits the impact of dataflow on timing errors, and does not need any hardware modification in the computing array.
Moreover, READ is a post-training optimization technique, so most algorithm-layer techniques are orthogonal to READ,
and READ can be also deployed on the timing speculation accelerator to reduce the toggle rate of TEDC modules, 
and enable more aggressive voltage scaling for higher energy efficiency.

\section{Motivation: Critical Pattern Reduction}
\label{sec:Motivation}
% To reduce the timing errors in the systolic array, 
We first analyze the factors affecting the TER in spatial accelerators.
For a specific cycle, the factors that determine whether a timing error will occur fall into two categories: 
1) the input pattern which determines the triggered paths; 
2) the operation conditions (including but not limited to temperature, voltage, and aging) that affect the path delay of these triggered paths.
The fluctuation of the operating conditions is often uncontrollable and only a small set of paths may exceed the clock cycle, i.e. critical paths.
Therefore, to reduce the timing error rate, we focus on reducing the trigger rate of the critical paths. 

The basic PE is the MAC unit.
We use the MAC unit in TPU as an example, which contains an 8-bit multiplier and a 24-bit accumulator.
Through performing dynamic timing analysis of the synthesized MAC unit, 
we find that the most common type of critical input pattern is the input that can cause the sign bit flip of partial sum (PSUM).
For example, when computing $3\times(-2) + 2 = -4$ in a certain MAC, the 2's complement representation is 
$00000011 \times 11111110 + 000000000000000000000010 = 111111111111111111111100$.
% the MAC operation $3\times(-2) + 2 = -4$, represented in 2's complement coding, would be $00000011\times11111110 + 000000000000000000000010 = 111111111111111111111100$.
% As can be observed, 
The flip of the sign bit triggers the long carry chain of the accumulator, and thus,
triggers the timing critical paths.
% It can be seen that the flip of the sign bit may trigger the long carry chain of the adder.
Similarly, it is also possible to trigger critical paths when PSUM changes from negative to positive.

\begin{figure}[tb]
    \centering
    \includegraphics[width=.4\textwidth]{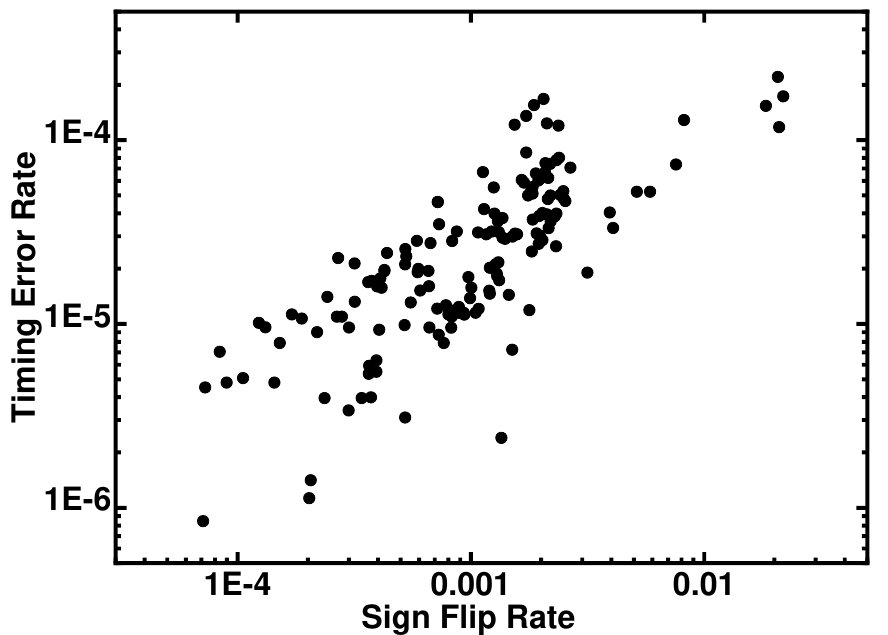}
    \caption{The sign flip rate and the timing error rate demonstrate a strong correlation.}
    \label{fig:ter-sfr}
    % \vspace{-0.3cm}
\end{figure}

To further validate this observation, we collect the sign flip rate and TER from different MAC units running different convolution layers with different dataflow.
As shown in Fig.~\ref{fig:ter-sfr}, the sign flip rate and the TER demonstrate a strong correlation, 
which indicates that most of the timing errors are caused by the sign flip of PSUM.
Hence, to minimize the TER, an effective approach is to reduce the sign flip rate of PSUM.

\section{Methodology}
\label{sec:Methodology}

In this section, we will formulate the problem of critical pattern reduction 
and propose a set of optimization techniques to reduce critical patterns and TER.
% We use the systolic array, the most widely used 2-D accelerator,
% with output stationary dataflow as an example to illustrate the algorithm.
% It should be noted that the algorithm proposed can be easily extended to other dataflow or 2-D spatial accelerators with minor modifications.
% Fig.~\ref{fig:sa-schematic} gives an example of mapping a 1-by-1 convolution to a systolic array with the output stationary dataflow.
% The weights are streamed along the row direction and the input activations are propagated in the column direction.
% Each MAC unit is responsible for all the computations required for a pixel in the output feature map.
We summarize the notations used in the paper in Table~\ref{table:notation}.
% The notations used in this paper are listed in Table~\ref{table:notation}.

\begin{table}[t]
\centering
\captionof{table}{Notations used in this paper.}
\label{table:notation}
\begin{tabular}{c|l}
\hline
$N,H,W,K$ & Output batch, height, width, channel        \\ \hline
$C,F_x,F_y$ & Input channel, filter height, filter width\\ \hline
$A_r, A_c$ & Rows and columns of the systolic array  \\ \hline
$W$    & Weight matrix                                  \\ \hline
$S$    & Sequence of input channel                      \\ \hline
$T$    & Cluster of output channel                      \\ \hline
\end{tabular}
% \vspace{-0.2cm}
\end{table}

\subsection{Problem Formulation}
\label{reorder-formulation}
PSUM is the intermediate accumulation result of a convolution operation, 
and the sign bit flip means that the sign bit of PSUM is different in two adjacent cycles.
The total number of sign bit flips computing an output activation is calculated as:
% \vspace{-0.3cm}
\begin{equation}
    SF = \sum_{j=0}^{C-1} \mathrm{sign}(\sum_{i=0}^{j}a_iw_i) \oplus \mathrm{sign}(\sum_{i=0}^{j+1}a_iw_i)
    \nonumber
    % \vspace{-0.2cm}
\end{equation}
% $$SF = \sum_{j=0}^{C-1} \mathrm{sign}(\sum_{i=0}^{j}a_iw_i) \oplus \mathrm{sign}(\sum_{i=0}^{j+1}a_iw_i)$$
where $\oplus$ denotes the XOR operation and $\mathrm{sign}(\cdot)$ extracts the sign bit of the input, which returns 1 when the input is positive and returns 0 for negative inputs.
% the sign function, which returns +1 when the input is positive and returns -1 for negative inputs.

\begin{figure}[!h]
    \centering
    \includegraphics[width=.48\textwidth]{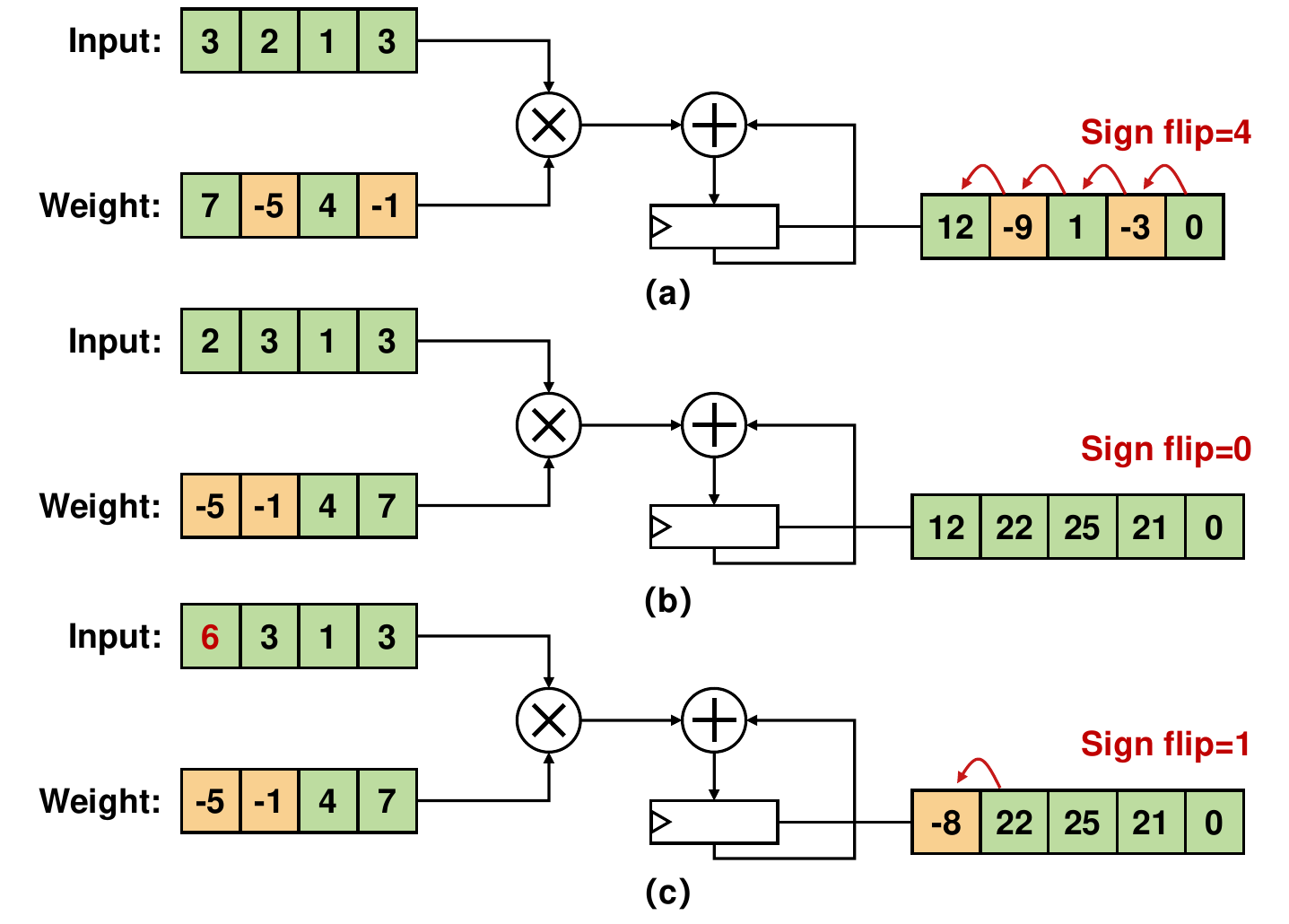}
    \caption{A 1$\times$4 convolution calculated in different orders. Reordering weights does not change the computing result, but avoids the critical input pattern of MAC.}
    \label{fig:reorder-schematic}
    % \vspace{-0.3cm}
\end{figure}

The optimization problem is combinatorial and the computation complexity scales exponentially, which quickly becomes intractable for even small-size problems. To efficiently solve the problem above, we make the following two observations:
\begin{itemize}
    \item As rectified linear unit (ReLU) is widely used in modern networks, the input activations $a_i$ of a convolution layer are often non-negative. Hence, the sign of each MAC result is mainly determined by the sign of weight $w_i$ and the sign flip of PSUM is mainly determined by the sequence of weights for the computation. Fig.~\ref{fig:reorder-schematic} gives a simple example to show how weight sequence impacts the sign flip.
    \item As the PSUM is usually initialized to 0, the minimum number of sign flips for when computing an output is 0 or 1, depending on whether the final activation is positive or negative.
\end{itemize}
Based on the observations above, we propose the following heuristic solution for the optimization above: \textit{when computing an output activation, arrange the computation sequence so that all MACs with non-negative weights are computed first}. The proposed solution has two main properties:
\begin{itemize}
    \item Compute correctness: as the convolution operation is independent of the computing order, changing the order of MAC operations only affects the value of PSUMs but not the final output activation.
    \item Sign flip optimality: Given non-negative inputs, after arranging the weights, the value of PSUM will first increase monotonically, and then monotonically decrease. If the final output activation is non-negative, PSUM is always non-negative, and no sign flip occurs (shown in Fig.~\ref{fig:reorder-schematic} (b)). If the final output activation is negative, PSUM will become negative and the sign flip occurs once (shown in Fig.~\ref{fig:reorder-schematic} (c)).
\end{itemize}
 % \vspace{-0.3cm}

\begin{figure}[tb]
    \centering
    \includegraphics[width=.49\textwidth]{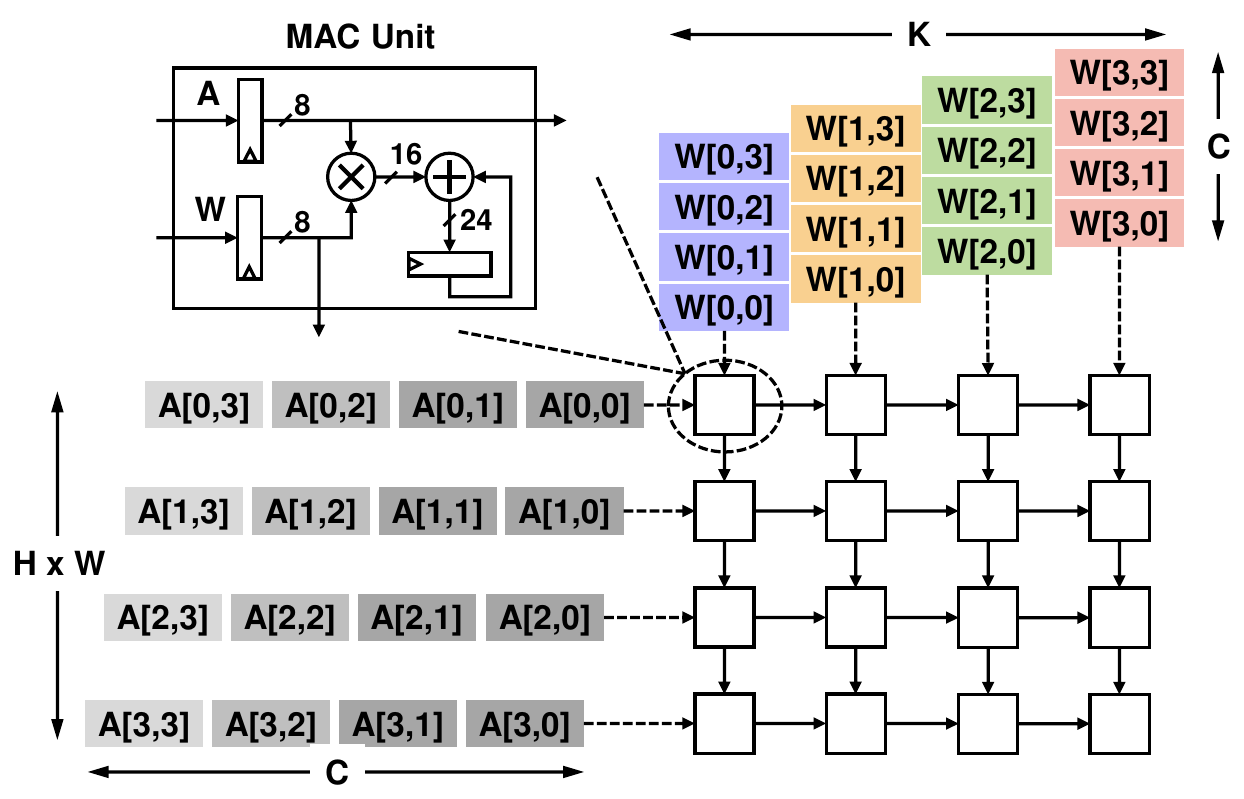}
    \caption{The architecture of a systolic array-based DNN accelerator with the output stationary dataflow.}
    \label{fig:sa-schematic}
    % \vspace{-0.3cm}
\end{figure}
% Because the commonly used activation function in convolution layer is rectifying linear unit (ReLU), the input activations $a_i$ are non-negative value.
% Therefore, the sign of each MAC result is mainly determined by the sign of weight $w_i$, which means that the sign flip of PSUM depends on the sequence of weights.
% \ml{The description below may make people believe our approach is too trivial. Need to re-write a bit to show how we come up with this heuristic method. Need to mention that due to ReLU, activation is always non-negative}
% One straightforward approach to achieve the minimum sign flip is to reorder the weights and input activations.
% Since the convolution operation is independent of the computing order, 
% changing the order of MAC operation only affects the value of PSUMs but not the final output activation.

% By arranging all non-negative weights to the front, 
% the value of PSUM will first increase monotonically, and then monotonically decrease.
% If the final output activation is non-negative, PSUM is always non-negative, and no sign flip occurs (shown in Fig~\ref{fig:reorder-schematic} (b)).
% If the final output activation is negative, sometime later in the calculation,
% PSUM will become negative and the sign flip occurs once (shown in Fig~\ref{fig:reorder-schematic} (c)).

\subsection{Input Channel Reordering}
% Inspired by the single convolution calculation example in Section~\ref{reorder-formulation}, 
% an effective technique to reduce the sign flip of a systolic array is to change the order the weight matrix streaming into the array.
% To find the optimal weight order, we define the input channel reordering problem as below.

% 晚些引入systolic array，免得审稿人以为我们的方法只能用于systolic array
We use the systolic array, the most widely used 2-D accelerator,
with output stationary dataflow as an example to illustrate the algorithm.
It should be noted that the algorithm proposed can be easily extended to other dataflow or 2-D spatial accelerators with minor modifications.

Fig.~\ref{fig:sa-schematic} gives an example of mapping a 1-by-1 convolution to a systolic array with the output stationary dataflow.
The weights are streamed along the row direction and the input activations are propagated in the column direction.
Each MAC unit is responsible for all the computations required for a pixel in the output feature map.

The heuristic solution proposed above can find the optimal sequence for the case of a single output channel.
However, to improve throughput and data reuse, systolic arrays often have more than one column to process multiple output channels simultaneously.
To reduce the sign flip of simultaneously processed channels, we need to find a sequence suitable for all these output channels. Hence, we define the input channels reordering problem as below.

\textbf{Problem 1} \textit{(Input Channel Reordering) Given a weight matrix $W \in \mathbb{R}^{C\times K}$, 
divide the $W$ into n sub-matrices $W_1,\dots,W_n$ $\in \mathbb{R}^{C \times A_c }$ according to the size of systolic array,
and find the optimal $S_1^*,\dots,S_n^*$ for each sub-matrix such that the sign flip of PSUM during the computing is minimized.
\footnote{We assume Fx = Fy = 1 for the weight matrix in formulation and algorithm, 
but the definition and analysis can be easily extended to cases where Fx and Fy are larger than 1.}}

% If we consider the case of single output channel, i.e., the sub-matrix $W_i \in \mathbb{R}^{C \times 1}$, 
% the reordering problem can be solved by sorting the input channels by the sign.
% However, to improve throughput and increase data reuse, systolic arrays often have more than one column to process multiple output channels simultaneously.
% So we need to sort the input channels for multiple output channels,
% in which weights in the same position may have different signs.

To efficiently solve the reordering problem, we propose two sorting algorithms as described in Algorithm\ref{algo:input_reorder}.
The first algorithm sorts the input channels according to the number of non-negative weights in each channel.
If two channels have the same number of non-negative weights, the channel with a larger sum of weights is ranked in the front.
We denote this method as the \textit{sign\_first} approach.
The Second algorithm sorts the input channels according to the sum of weights in each channel.
If the sums of two channels are the same, the channel that has more non-negative weights is ranked in front.
We denote this method as the \textit{mag\_first} approach.

\begin{algorithm}[tb]
    \caption{Input channel reorder.}
    \label{algo:input_reorder}
    \SetKwProg{Fn}{Function}{:}{end}
    
    \KwIn{Weight matrix $W$}
    \KwOut{Sequence $S$ of each sub-matrix}
    \Fn{sort\_input\_channel ($matrix\ W$)} {
        \For{$i=1:C$}{
            $metric\_sign[i] = \sum_{j=1}^{A_c}sign(W[i,j])$\;
            $metric\_mag[i] = \sum_{j=1}^{A_c}W[i,j]$\;
        }
        
        \If{$sorting\_criteria == "sign\_first"$}{
            % $scale\ metric\_mag\ to\ between\ 0\ to\ 1$\;
            scale $metric\_mag$ to between 0 to 1\;
        }\Else{
            % $scale\ metric\_sign\ to\ between\ 0\ to\ 1$\;
            scale $metric\_sign$ to between 0 to 1\;
        }
            $metric\_sorting = metric\_sign + metric\_mag$\;
            $S = argsort(metric\_sorting)$\;
        \Return S\;
    }
    \BlankLine
    Divide $W$ into $n$ sub-matrix $W_1,\dots,W_n$\;
    \For{$i = 1:n$}{
        $S_i = sort\_input\_channel(W_i)$\;
        reorder sub-matrix $W_i$ in order $S_i$\;
        reorder input activations in order $S_i$\;
    }
\end{algorithm}

\begin{figure*}[ht]
    \centering
    \subfloat[] {\includegraphics[width=.25\textwidth]{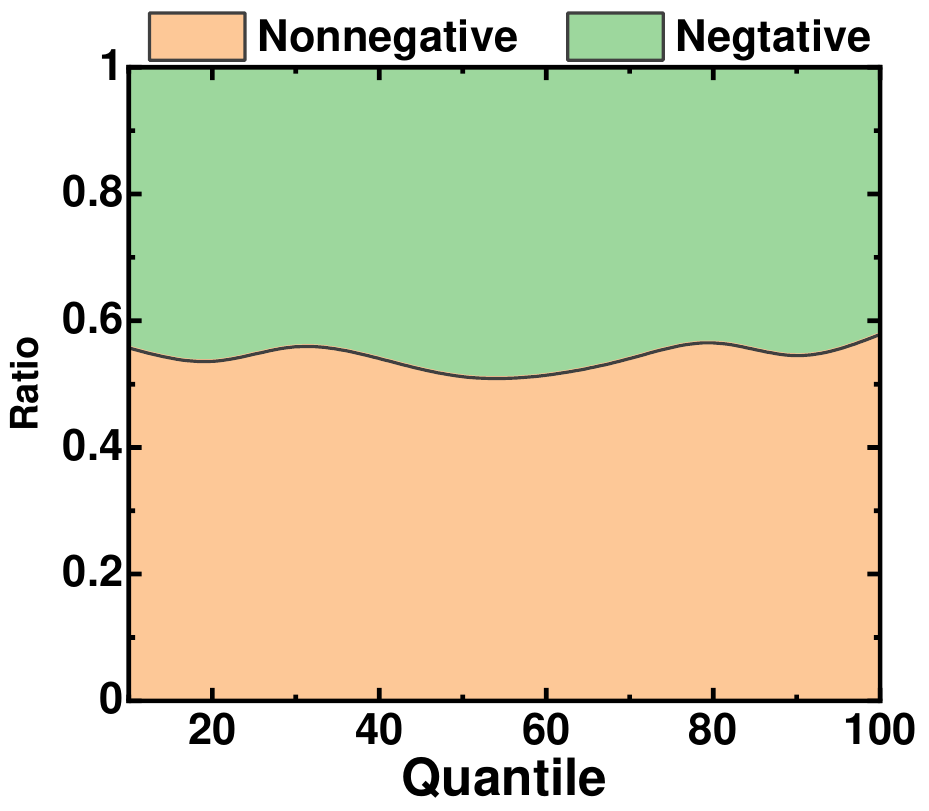}}
    \subfloat[] {\includegraphics[width=.25\textwidth]{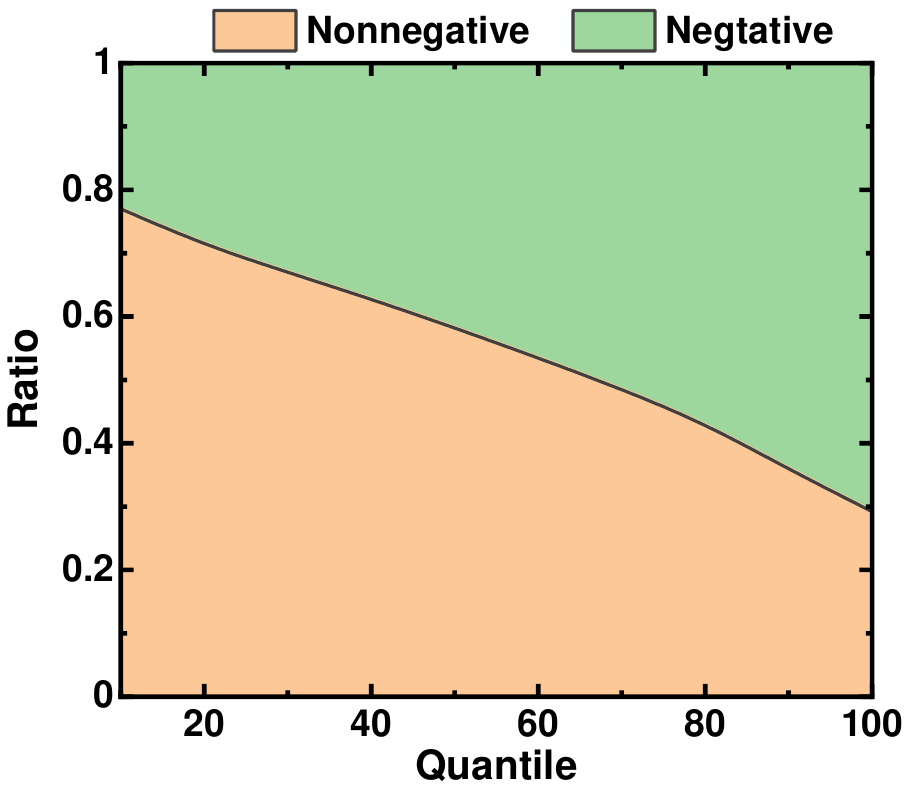}}
    \subfloat[] {\includegraphics[width=.25\textwidth]{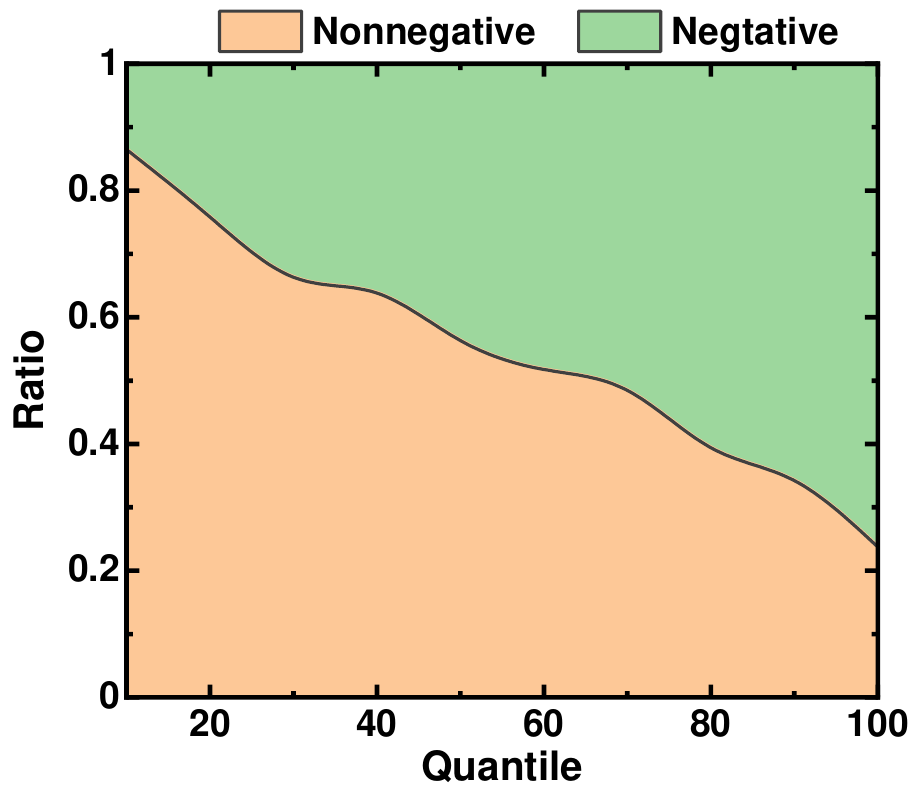}}
    \subfloat[] {\includegraphics[width=.25\textwidth]{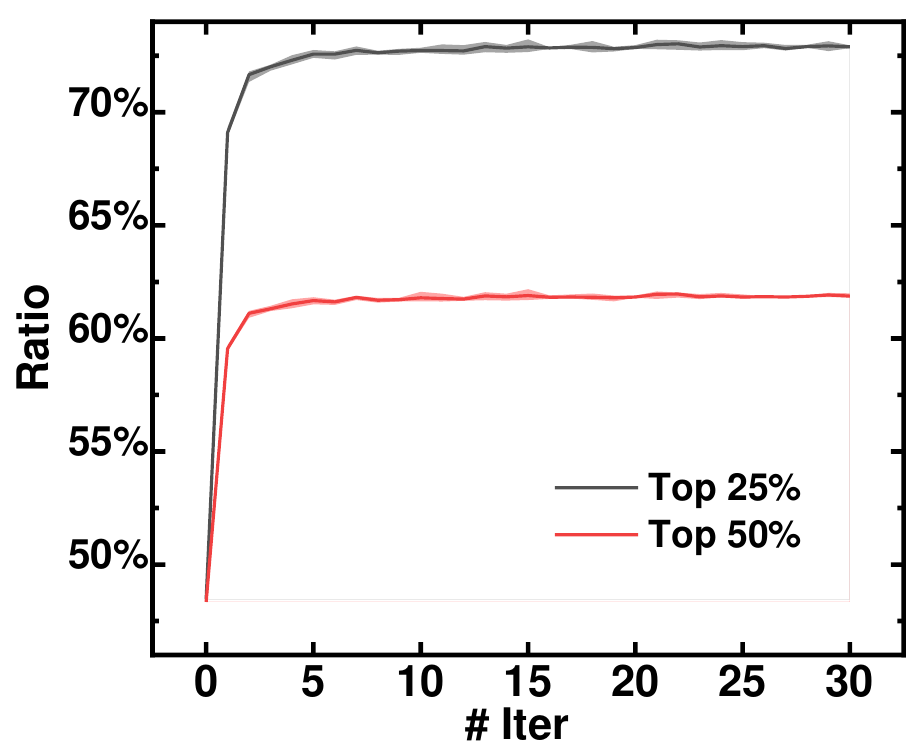}}
    % \subfloat[] {\includegraphics[width=.25\textwidth]{fig/fig4-ter-col.pdf}}
    \caption{
    The proportion of negative and non-negative weights in different positions of 
    (a) the initial weight matrix, (b) the weight matrix reordered by the \textit{mag\_first} approach, and (c) the weight matrix reordered by the \textit{sign\_first} approach.
    (d) The ratio of non-negative weights in the top 25\% and 50\% of the weight matrix.
    % (c) Output channel clustering further concentrates the non-negative weights into the font of the matrix.
    % (d) TER with different reordering algorithms for different channels per cluster.
    }
    \label{fig:reorder}
    % \vspace{-0.3cm}
\end{figure*}

We use a convolution layer of the VGG-16 on the Cifar10 dataset as an example and run the two reordering algorithms.
Fig.~\ref{fig:reorder} (a) shows the proportion of negative and non-negative weights in different positions of the initial weight matrix.
It can be seen that the distribution of non-negative and negative weights is uniform.
Fig.~\ref{fig:reorder} (b) and (c) show the proportion for the reordered weight matrix, 
in which the non-negative weights concentrate in the front and the negative weights are mainly distributed in the back.
And the reordering results of the \textit{sign\_first} approach are better.

% We then perform the dynamic timing analysis to estimate the TER by assuming 10-year aging and 5\% voltage and temperature variations
% (see Section~\ref{sec:setup} for detailed experimental setup).
% The results are shown in Fig.~\ref{fig:reorder} (d), 
% although both reordering algorithms become less effective as the number of array columns $A_c$ increases, 
% there is a significant reduction in the TER compared to the baseline without reordering.
% Moreover, the \textit{sign\_first} approach achieves larger TER reduction compared to \textit{mag\_first}, probably because most of the weights are small.
% Therefore, unless specifically mentioned in the rest of paper, the reordering algorithm used are the \textit{sign\_first} approach.
% \vspace{-0.2cm}
\subsection{Output Channel Clustering}
% The reason why the effectiveness of reordering is affected by the number of columns $A_c$
% is that the location of the negative weights varies across different output channels, 
% hence it is difficult to find an order that is suitable for all channels.

To further improve the effectiveness when the number of columns $A_c$ is large, 
we propose to cluster the output channels first before segmenting the weight matrix.
Then, the input channels are reordered for each cluster separately.
We denote this approach as \textit{cluster-then-reorder}. 
Consider the example of the following weight matrix:

\begin{small}
\begin{equation}
 W = \left[
    \begin{array}{cccc}
        \tikzmarkin[lightgray]{col 1} 4 & -5 &  \tikzmarkin[lightgray]{col 2} 5 & -1\\
        -10 &  3 & -2 & 2\\
          9 & -2 &  3 & -1\\
         -2 \tikzmarkend{col 1}&  3 & -6 \tikzmarkend{col 2}& 3\\
    \end{array}
\right]
\nonumber
\end{equation}
\end{small}
Assume the computing array has 2 columns and only allows for streaming 2 output channels simultaneously. 
Instead of directly segmenting $W$, we can first cluster the output channels into 2 groups, i.e., {0, 2} and {1, 3}, 
and then segment $W$ into $W_1$ and $W_2$ as below:

\begin{small}
\begin{equation}
 W_1 =\left[
    \begin{array}{cc}
       \tikzmarkin[lightgray]{col 3}4 & \tikzmarkin[lightgray]{col 4}5 \\
        -10 & -2 \\
        9   &  3 \\
        -2 \tikzmarkend{col 3} & -6 \tikzmarkend{col 4}\\
    \end{array}
\right], 
W_2 = \left[
    \begin{array}{cc}
        -5 & -1\\
         3 &  2\\
        -2 & -1\\
         3 &  3\\
    \end{array}
\right]
\nonumber
\end{equation}
\end{small}
Compared to direct segmenting, we can assign output channels with similar sign positions to the same cluster, which makes it easier to reorder each sub-matrix.
% the clustering makes the sorted sub-matrix more ordered, 
% because output channels with similar sign positions are assigned to the same cluster.

To formulate the clustering problem,
we first define the sign difference (SD) between two $n$-dimensional vectors $x$ and $y$ as the Manhattan distance of two vectors:
% \begin{equation}
%     \begin{aligned}
%         SD(x,y) &= D_{m}(sign(x),sign(y)) 
%                 &= \sum_{i=1}^{n} \lvert sign(x[i]) - sign(y[i]) \rvert 
%     \end{aligned}
%     \nonumber
% \end{equation}
% \vspace{-0.2cm}
\begin{equation}
    SD(x,y) = \sum_{i=1}^{n} \lvert sign(x[i]) - sign(y[i]) \rvert 
    \nonumber
    % \vspace{-0.2cm}
\end{equation}
% where $D_m(\cdot)$ is the function that calculates the Manhattan distance of two vectors.
Let ${T_1,\dots,T_n}$ denote the $n$ clusters of output channels, and $W_{T_1},\dots,W_{T_n}$ denote the corresponding $n$ sub-matrices.
Similarly, the SD of a sub-matrix $W_{T_i}$ can be defined as

% \vspace{-0.2cm}
\begin{equation}
        SD(W_{T_i}) = \sum_{i,j \in T_i}SD(W_{T_i}[:,i], W_{T_i}[:,j])
    \nonumber
    % \vspace{-0.2cm}
\end{equation}

The smaller the SD is for a sub-matrix, the easier it is to reorder the input channels and reduce sign bit flips. Hence, the output channel clustering problem can be defined as follows.

\textbf{Problem 2} \textit{(Output Channel Reordering) Given a weight matrix $W \in \mathbb{R}^{C \times K}$, 
find $n$ clusters ${T_1,\dots,T_n}$ such that the sign difference of each sub-matrix $W_{T_i}$ is minimized, i.e.,}
% \vspace{-0.3cm}
\begin{equation}
    \begin{aligned}
    \min_{T_1,\dots, T_n} \quad & \sum_{i=1}^{n} SD(W_{T_i})\\
    s.t. \quad & T_i \cap T_j = \emptyset
    \quad \forall i \neq j \\
    & \cap_{i=1}^n T_i = \{1,\dots,K\}
    \end{aligned}
    \nonumber
\end{equation}
% \vspace{-0.3cm}

This is a hard-balanced clustering problem, which can be solved by many proven algorithms.
% We adopt the balanced KNN~\cite{balanced_knn} on the weight sign matrix by the Manhattan metric to solve this problem.
We adopt the balanced KNN on the weight sign matrix by the Manhattan metric to solve this problem.

The convergence plot of the clustering algorithm is shown in Fig.~\ref{fig:reorder} (d).
We found that the clustering algorithm further concentrates the non-negative weights into the front of the matrix and converges very well.

\subsection{Hardware Support}
%\ml{Need to update here since special hardware support is needed.}

\begin{figure}[tb]
    \centering
    \includegraphics[width=.42\textwidth]{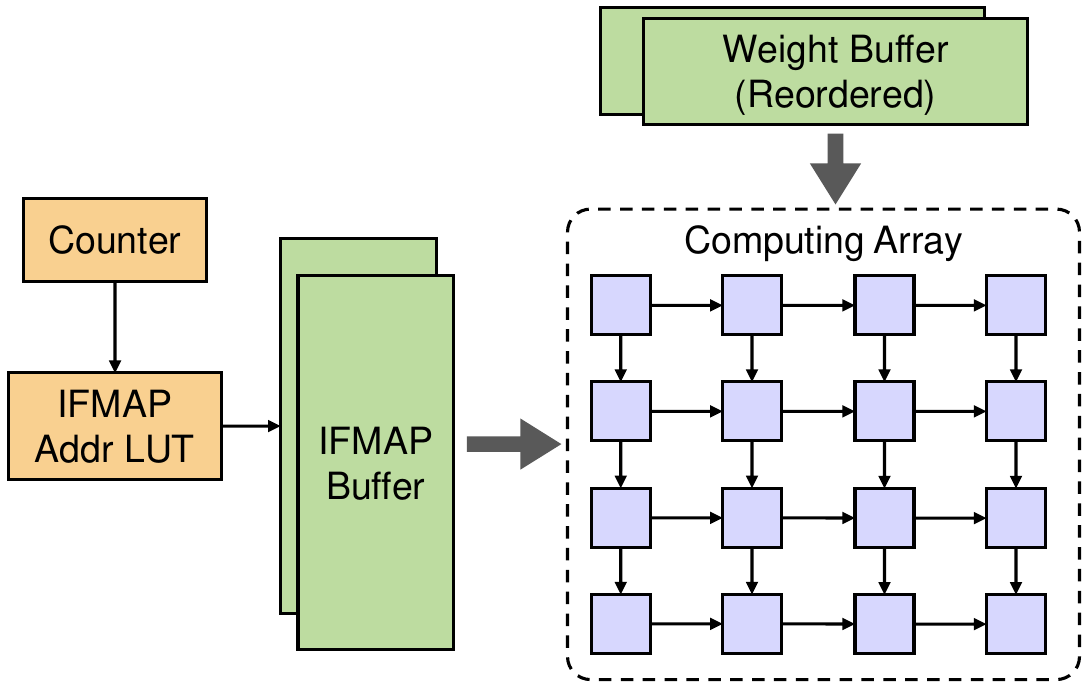}
    \caption{Hardware support for the proposed technique.}
    \label{fig:hardware-modi}
    % \vspace{-0.3cm}
\end{figure}

The input channel reordering algorithm switches the weight matrices and input sequences. 
While the weight matrices can be reordered before inference, the input activations have to be reordered during inference. 
This is because we use different input sequences for different clusters of output channels. 
As shown in Fig~\ref{fig:hardware-modi},
the reordering can be realized by simply augmenting the activation read logic with an address lookup table (LUT) to enable accessing reordered activations.
For a layer with 1024 channels, the required LUT SRAM size is less than 2KB. 
Compared to the original on-chip buffer (usually 2$\sim$64MB), the energy and area overhead of LUT are negligible.

% The input channel reordering algorithm switches the input sequence, so the weight matrix need to be reordered in memory before inference,
% and the reordering of input activations requires an address lookup table (LUT) to generate the actual address. 
% For a reasonable buffer depth, i.e., 1024, the required LUT SRAM size is less than 2KB, which has negligible energy and area overhead.

Output channel clustering will change the order of output activations, hence it impacts the memory fetching of the next layer. 
As the reordering algorithm across layers proposed in~\cite{channel_permutation},
starting from the second layer, memory fetching of input activations of each layer is determined by two orders:
the weight order of the current layer, applied along its $C$ dimension, and the output channel order of the previous layer, applied along its $K$ dimension.

The proposed algorithm only changes the order of computing, without introducing any additional computation. 
Moreover, as the computing order is determined before inference, the process of writing to LUT SRAM can be implemented concurrently with the 
writing of weights and input activations, without affecting the throughput. 
Additionally, as the size of the LUT SRAM is small, the impact on bandwidth is negligible.
\section{Experimental Results}
\label{sec:Results}

\begin{figure}[tb]
    \centering
    \includegraphics[width=.38\textwidth]{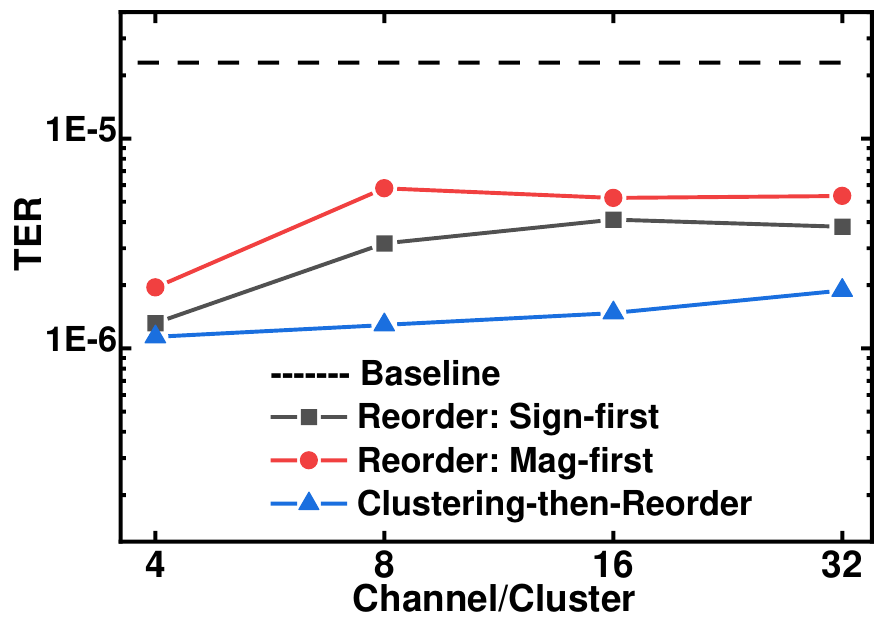}
    \caption{TER with different reordering algorithms for different channels per cluster.}
    \label{fig:ter-col}
    % \vspace{-0.3cm}
\end{figure}

\subsection{Experimental Setup}
\label{sec:setup}
We use VGG-16 and ResNet-18 trained on the CIFAR10 and CIFAR100 dataset, and pre-trained ResNet-34 on ImageNet for evaluation.
We designed an output-stationary systolic array with 16 rows and 4 columns.
Each MAC unit in the array can support the multiplication and accumulation of 8-bit activations, 8-bit weights, and 24-bit partial sums.
The array is synthesized with the open-source Nangate 15nm standard cell library~\cite{nangate}, using Synopsys Design Compiler.
The nominal frequency is determined by the static timing analysis with Synopsys Primetime.
TER of MAC units under PVTA variations is evaluated by the dynamic timing analysis framework proposed in~\cite{avatar}.
For voltage and temperature fluctuation, we use Siliconsmart to generate the LVF libraries at nominal, 3\%, and 5\% VT fluctuation with a commercial 16/14 nm FinFET modelcard. 
For aging analysis, we consider the negative bias temperature instability (NBTI) as NBTI dominates the transistor aging in digital circuits~\cite{Guo_nbti}.
The baseline results are from the same architecture as the original computation sequence.

\begin{figure*}[tb]
    \centering
    \includegraphics[width=.95\textwidth]{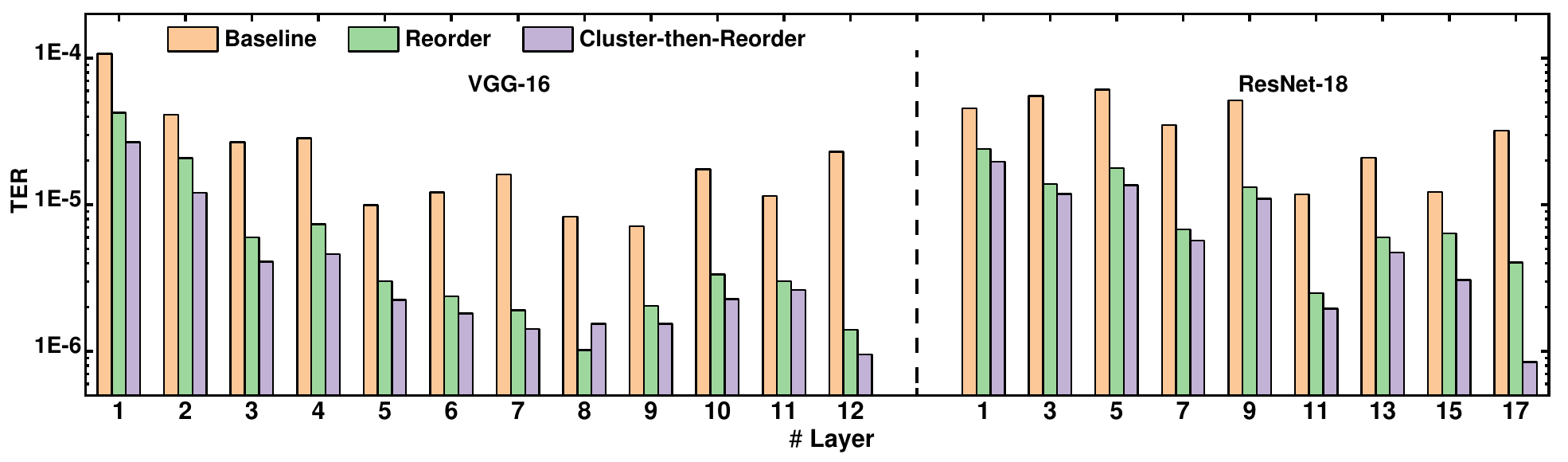}
    \caption{Timing error rate comparison for different reliability-enhanced algorithms on ResNet-18 and VGG-16.}
    \label{fig:ter-layer}
    % \vspace{-0.3cm}
\end{figure*}

\subsection{Layer-wise TER Reduction}
We first compare the layer-wise TER of different algorithms.
The TER is estimated by assuming 10-year aging and 5\% voltage and temperature fluctuation.
The results are shown in Fig.~\ref{fig:ter-col}, 
although reordering algorithms become less effective as the number of array columns $A_c$ increases, 
there is a significant reduction in the TER compared to the baseline without reordering.
Moreover, the \textit{sign\_first} approach achieves a larger TER reduction compared to \textit{mag\_first}, probably because most of the weights are small.
Compared to direct reordering, the \textit{cluster-then-reorder} algorithm performs better, especially when the number of columns $A_c$ is relatively large.
Therefore, unless specifically mentioned in the rest of the paper, the reordering algorithm used is the \textit{sign\_first} approach.

% We use \textit{sign\_first} criteria to sort the input channels for both algorithms.
As shown in Fig.~\ref{fig:ter-layer}, the average TER reduction of the direct reordering and the cluster-then-reorder algorithms are 4.9$\times$ and 7.8$\times$, respectively.
% is 4.9$\times$ and the average TER reduction of the cluster-then-reorder algorithm is 7.8$\times$.
The \textit{cluster-then-reorder} algorithm usually results in a higher reduction for most of the layers, and it has better performance in the later layers of the network since the number of output channels is larger in the later layers.

\begin{figure}[tb]
    \centering
    \includegraphics[width=.48\textwidth]{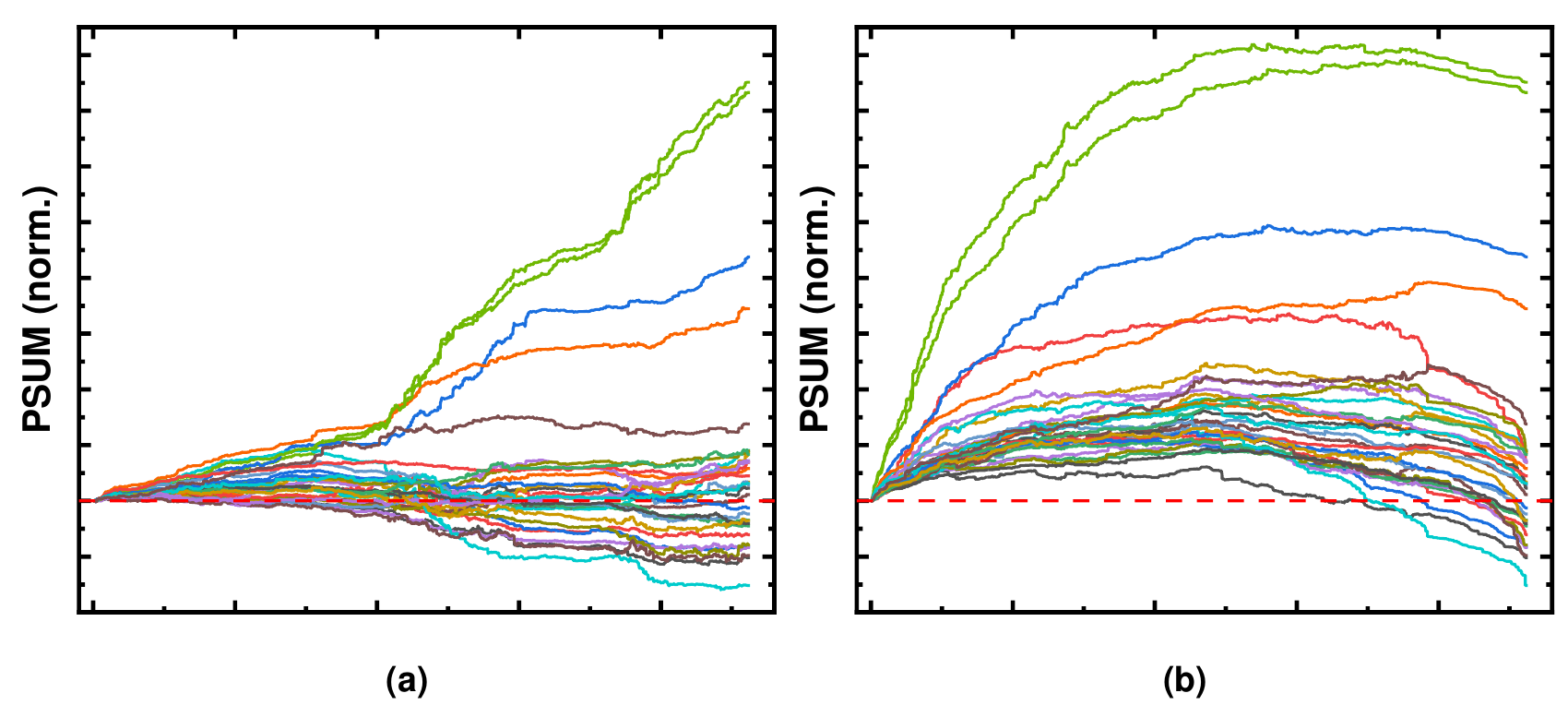}
    \caption{The accumulation of PSUM in convolution in (a) original sequence and (b) reordered sequence.}
    \label{fig:psum_trend}
    % \vspace{-0.3cm}
\end{figure}

Fig.~\ref{fig:psum_trend} gives a fine-grained view to show how reordering can reduce timing errors.
We show the accumulation of PSUM during multiple convolutions on a MAC when running VGG-16.
Because the algorithm puts the positive weights in the front, 
the PSUM will increase first and then decrease (Fig.~\ref{fig:psum_trend} (b)),
which significantly reduces the number of sign flips (through the red dash line).
Fig.~\ref{fig:psum_trend} also indicates that by adopting the proposed reorder technique,
the sign flip rate of PSUM for a specific NN layer is also determined by the proportion of the negative value in output activations of the current layer.
The smaller the proportion of negative output activations, the lower timing error rate.

Moreover, we observed that the TER reduction of a certain layer is related to the distribution and sparsity of the weights. 
Layers with a higher proportion of non-negative weights tend to have more ordered sorting weights, and result in better TER reduction.
We also found that weight matrices with higher sparsity tend to cluster and sort more easily. 
The proposed method did not perform well due to high proportions of negative weights in certain layers.
But the proposed method achieved a significant reduction in error rate across all tested layers.
These results suggest that the TER can be further improved by adjusting the weight matrix according to certain rules during training.

\begin{figure}[tb]
    \centering
    \includegraphics[width=.45\textwidth]{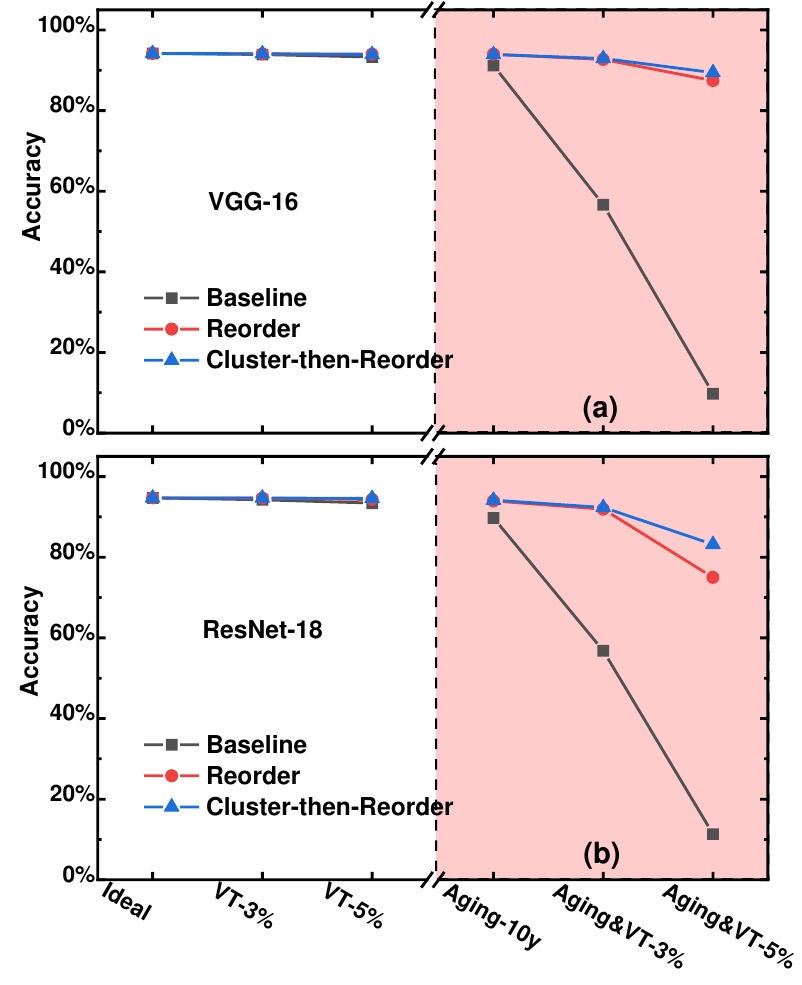}
    \caption{Accuracy of (a) VGG-16 and (b) ResNet-18 on Cifar10 dataset under various PVTA conditions with different optimization techniques.}
    \label{fig:nn-accuracy}
    % \vspace{-0.3cm}
\end{figure}

\subsection{Accuracy Improvement}
After obtaining the layer-wise TERs, we use the equation~\ref{BER_cal} to calculate the BERs of the output activations.
Then, we perform an error-injection simulation to evaluate the inference accuracy~\cite{ares_dnn}.
The error-injection simulation is implemented using PyTorch.
We randomly flip the corresponding bits of the output activations (before the activation function) according to the BERs.
To avoid randomness error, the batch size is set to 128 and the error-injection simulation is repeated five times with different seeds for each BER combination.

\begin{figure}[tb]
    \centering
    \includegraphics[width=.45\textwidth]{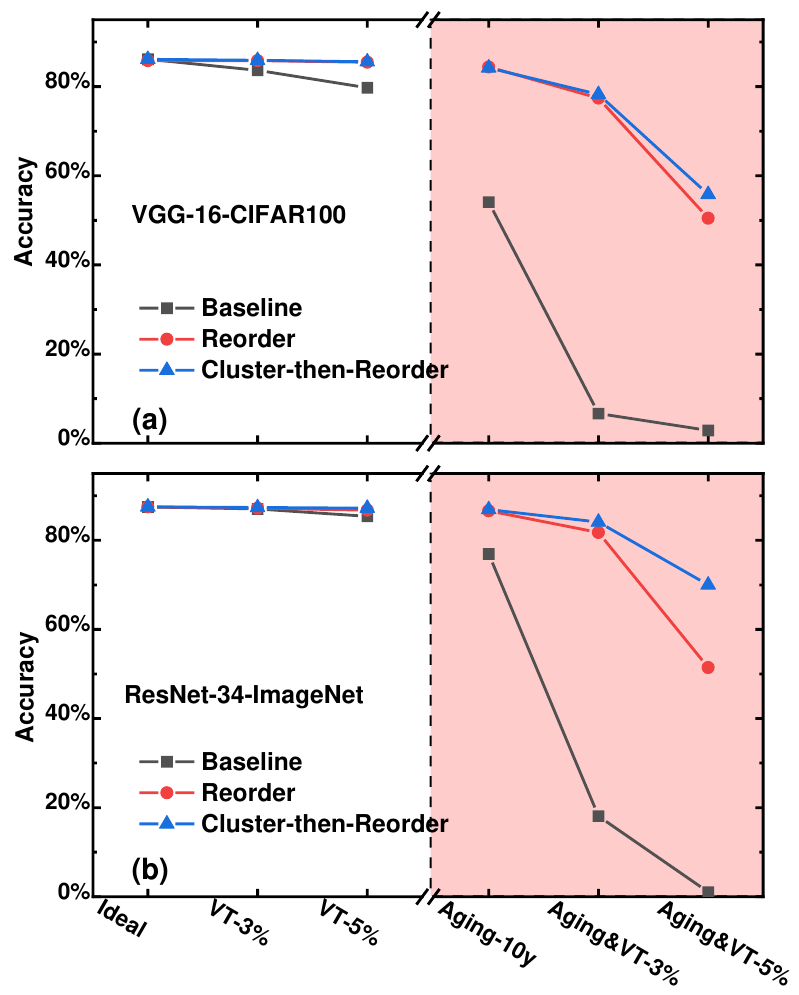}
    \caption{Top-3 accuracy of (a) VGG-16 on CIFAR100 and (b) ResNet-34 on ImageNet under various PVTA conditions with different optimization techniques.}
    \label{fig:nn-accuracy2}
    % \vspace{-0.3cm}
\end{figure}

Fig.~\ref{fig:nn-accuracy} shows the accuracy of VGG-16 and ResNet-18 on CIFAR10 under various PVTA conditions with different optimization algorithms.
The accuracy loss of the baseline setting without any optimization is significant with the PVTA variations, especially after 10-year aging.
While the accuracy of the proposed reordering algorithm and cluster-then-reorder algorithm is kept in an acceptable range.

Fig.~\ref{fig:nn-accuracy2} shows the top-3 accuracy of VGG-16 on CIFAR100 and ResNet-34 on ImageNet under various PVTA conditions.
The trend is similar to Fig.~\ref{fig:nn-accuracy}.
To speed up the simulation, we injected errors only into several vulnerable layers (those closer to the inputs).
Although the proposed algorithm shows accuracy loss under extreme fluctuations, it still withstood a wider range of fluctuations compared to the baseline.

The proposed reordering algorithm can be adopted not only to improve the robustness of accelerators against PVTA variations, 
but also to optimize power consumption.
A promising technique for low-power accelerators, timing speculation, uses Razor flip-flops for detecting and recovering timing errors.
However, the detection and correction of timing errors by Razor flip-flops also cause significant power consumption. 
Without an appropriate design methodology, timing speculation can lead to additional power consumption that negates much of the energy efficiency gains achieved by operating at lower voltages.
The proposed reordering algorithm in this paper can also reduce the toggle rate of Razor flip-flops in timing speculation accelerators, 
and allow for more aggressive voltage scaling while maintaining acceptable inference accuracy, resulting in higher energy efficiency.

% \begin{figure*}[tb]
%     \centering
%     \includegraphics[width=.80\textwidth]{fig/fig6-accuracy.pdf}
%     \caption{Accuracy of two benchmark DNN networks under various PVTA conditions with different reliability optimization methods.}
%     \label{fig:nn-accuracy}
% \end{figure*}

% \begin{figure}[tb]
%     \centering
%     \subfloat[] {\includegraphics[width=.45\textwidth]{fig/fig6-1-accuracy-dotline.pdf}}
%     \quad
%     \subfloat[] {\includegraphics[width=.45\textwidth]{fig/fig6-2-accuracy-dotline.pdf}}
%     \caption{Accuracy of (a) VGG-16 and (b) ResNet-18 under various PVTA conditions with different optimization techniques.}
%     \label{fig:nn-accuracy}
% \end{figure}

% \input{doc/5-related-works}
\section{Conclusion}
\label{sec:Conclusion}

In this paper, we present READ, a reliability-enhanced accelerator dataflow optimization algorithm.
With the proposed output channels clustering and input channels reordering algorithm, the TER of the datapath can be significantly reduced.
Evaluation with VGG and ResNet neural networks shows that the proposed methods can reduce TER by an average of 7.8$\times$ and up to 37.9$\times$, 
and make the accelerator more robust to PVTA variations, which demonstrates significant potential in reliable and efficient neural network accelerator design.
\section*{Acknowledgements}

This work was supported in part by the NSFC (62125401, T2293700, T2293704) and the 111 Project (B18001)

\bibliographystyle{IEEEtran}
\bibliography{./ref/Top.bib, ./ref/Software, ./ref/dta}

\end{document}